\documentclass[12pt]{article}
\usepackage{a4,latexsym,pslatex,graphicx,times,color}

\usepackage[numbers]{natbib}

\title{Bubble entrainment by a sphere falling through a horizontal soap foam}
\author{S.J. Cox, I.T. Davies \\ 
Department of Mathematics, Aberystwyth University,\\
Aberystwyth SY23 3BZ, UK.}
\date{}

\begin{document}

\maketitle

\begin{abstract}
We simulate the quasi-static motion of a spherical particle through a stable, horizontal soap film. The soap film subtends a fixed contact angle, in the range $10-135^\circ$, where it meets the particle. The tension and pressure forces acting on the particle are calculated in two cases: when the film is held within a vertical cylinder, trapping a bubble but otherwise free to move vertically, and when the outer rim of the film is held in a fixed circular wire frame. Film deformation is greater in the second case, and the duration of the interaction therefore increases, increasing the contact time between particle and film. As the soap film returns towards its equilibrium shape following the passage of the particle a small bubble is trapped for contact angles below a threshold value of $90^\circ$. We show that this bubble is larger when the particle is larger and when the contact angle is smaller. 
\end{abstract}

\section{Introduction}
\label{sec:intro}

Aqueous foams interact with particles in a number of important situations~\cite{WeaireH99,mousse13}; at high particle density the particles can even replace surfactant and stabilise the foam~\cite{binksM06}. At the other extreme, foam films can be used to separate individual particles based on their size~\cite{stogingfcww18}. In between, processes such as froth flotation and explosion suppression~\cite{mousse13,monloubou16} rely on the extent to which particles are trapped by foam films. Once in the film, particles may rotate and, depending on parameters such as the contact angle, may cause rupture~\cite{morrisnc12}.

\citeauthor{legoffcsq08}~\cite{legoffcsq08} found that small millimetric-sized particles falling on to a soap film at speeds of about 1 {\rm m/s} do not break the film. That is, after the particle has passed through the soap film the film ``heals'' itself \cite{courbins06}. This arrangement of a stable soap film held horizontally while a small spherical particle falls onto it permits an investigation of the forces that the soap film exerts on the particle and the consequent changes to the particle's velocity. The soap film can be considered to represent one repeating unit of a more extensive ``bamboo" foam~\cite{daviesc12}, in which successive impacts between the particle and different soap films could bring the particle to rest, representing a microscopic approach to the way in which a foam can be used in impact protection~\cite{monloubou16}. In the following, we choose the particle's weight sufficiently large that it is never trapped by a single soap film. Then the film is pulled into a catenoid-like shape as it is stretched by the particle, until, similar to the usual catenoid instability \cite{cryers92}, the neck collapses and the soap film returns to its horizontal state.

We will show that the forces exerted on the particle depend strongly on the contact angle along the triple line (Plateau border) where the liquid, gas and solid particle meet. In an experiment this contact angle could be adjusted by coating the particle~\cite{teixeiraact18}. We allow the contact angle at which the soap film meets the spherical particle to vary: the equilibrium case is a contact angle of $\theta_c = 90^\circ$~\cite{davies18}, in which the sphere is assumed to be coated with a wetting film that allows the soap film to move freely.  However, experimental photographs~\cite{legoffcsq08,chenpjhd19} show that the soap film wraps around the particle, with a contact angle far from $90^\circ$, before forming a catenoid-like neck. This suggests that the particle's motion is faster than the mechanical relaxation of the foam. Here we nonetheless employ quasistatic simulations, and presume that the only effect of the dynamic nature of the experiments is to adjust the contact angle between particle and film. We consider several values of $\theta_c$ down to $10^\circ$. 

In experiments, the collapse of the catenoidal neck above the particle generates a small bubble~\cite{legoffcsq08}, as for the impact of a liquid drop on a liquid surface~\cite{oguz_prosperetti_1990,thoroddsenteh05} and the collapse of an isolated soap-film catenoid~\cite{robinsons01}. This small bubble was not seen in previous simulations with a $90^\circ$ contact angle~\cite{daviesc12,davies18}. Our new simulations make clear why this is the case: only with a contact angle smaller than $90^\circ$ does the film curve around the particle sufficiently before detachment to enclose such a volume of gas.

The particle in our simulations, described below, is a sphere of given radius $R_s$ and mass $m$ grams, and hence with density $\rho = m / (4/3 \pi R_s ^3)$. It falls, initially under its own weight, towards a film with interfacial tension $\gamma = 30 {\rm mN/s}$ (so a film tension of $2\gamma$). We consider two cases: 
\begin{enumerate}
\item the soap film is held in a cylinder of radius $R_{cyl} = 1{\rm cm}$ and height $H = 2 {\rm cm}$. The film encloses a bubble of fixed volume  $0.5 H \pi R_{cyl}^2 = 0.5\pi {\rm cm}^3$, i.e. that fills the lower half of the cylinder. In this case both the tension in the film and the pressure in the bubble exert a force on the sphere once it touches the film.
\item the soap film is held by a fixed ring of radius $R_{cyl} = 1{\rm cm}$. In this case only the tension in the film exerts a force on the sphere.
\end{enumerate}

The Bond Number is defined as $Bo = \frac{1}{2} \rho g R_s^2 / \gamma$. In the simulations in \S\ref{sec:results} we ensure that the Bond number is just greater than one, indicating that gravitational forces should exceed the retarding force due to surface tension. Making the density (and hence the Bond number) smaller would lead to the sphere being trapped by the film, while increasing it would mean that the quasistatic approximation that we employ would be less appropriate. The maximum vertical tension force that the soap film could exert on the sphere to counteract its weight occurs when the film meets the sphere on its equator and pulls vertically upwards; then the film tension multiplied by the sphere circumference is $4 \pi \gamma R_s$. So, roughly speaking, if the particle density is constant then to pass through the film requires that the particle radius must be greater than $\sqrt{3 \gamma / (\rho g) }$ whereas if the particle mass is constant the condition is $R_s < mg/(4 \pi \gamma)$.

\section{Method}

\begin{figure}
\centerline{
 \includegraphics[angle=0,width=0.65\textwidth]{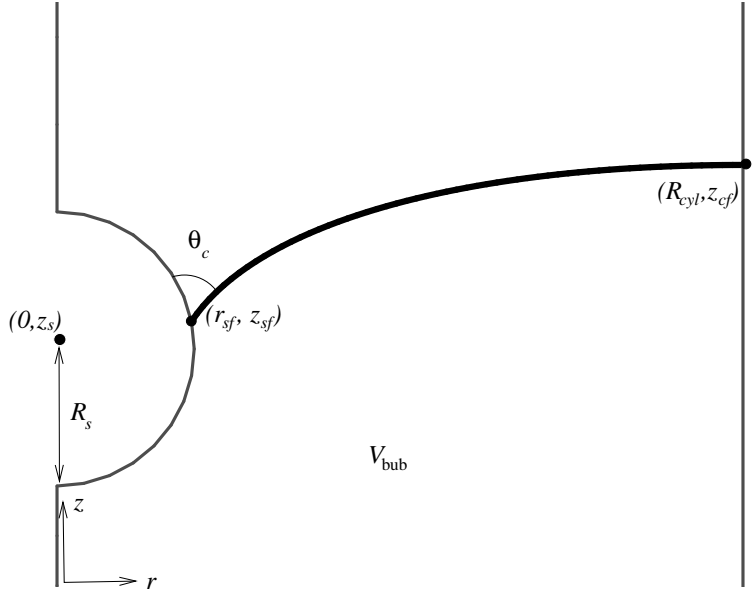}
  }
\caption{The axisymmetric structure under consideration, shown in the $(r,z)$ plane. In case 1 there is a bubble of fixed volume $V_{bub}$ and the vertex at position $(R_{cyl}, z_{cf})$ is free to move,  while in case 2 there is no volume constraint and the vertex  is fixed.}
\label{fig:notation}
\end{figure}

\subsection{Geometry}

We use the Surface Evolver~\cite{brakke92} to compute the shape of the soap film. Since this software gives information about static situations, we assume that the motion is overdamped, and therefore that the sphere and soap film move through a sequence of equilibrium positions determined by the forces acting.

By symmetry the sphere must remain in the centre of the film, so we perform an axisymmetric calculation in the $(r,z)$ plane. See figure~\ref{fig:notation}. The film is represented by a curve whose endpoints touch, respectively, the sphere (or the axis of the cylinder before attachment and after detachment) and the outer cylinder / ring. We discretize the curve into short straight segments of length ${\rm d}l$ and write the energy of the system as
\begin{equation}
 E_{film} = 2 \gamma \sum_{segments} 2 \pi r {\rm d }l.
\end{equation}
We restrict segments to have lengths in the range 0.01 to 0.05 which balance the need for accuracy with short computational time.

To include a contact angle $\theta_c$ we add a further term to the energy representing a spherical cap of film with tension $2 \gamma \cos \theta_c$ that covers the lower part of the sphere. This is based on the height $z_{sf}$ of the film where it meets the sphere:
\begin{equation}
 E_{\theta_c} = 2 \gamma \cos \theta_c \; . \;  2 \pi R_s \left(z_{sf} - (z_s-R_s) \right),
\end{equation}
where $z_s$ is the height of the centre of the sphere. This energy is set to zero before attachment and after detachment.

In case 1 we must also account for the volume $V_{bub}$ of the bubble trapped beneath the soap film. We calculate this volume based on the shape of the film and the positions of its endpoints. There are three terms required: 
\begin{eqnarray}
V_1 & = & \sum_{segments} \pi r^2 {\rm d }z \nonumber \\
V_2 & = & \left\{ 
	\begin{array}{ll}
	 0 & z_{sf} < z_s - R_s\\
	  \pi R_s^2 (z_{sf}-z_s)-\frac{\pi}{3}(z_{sf}-z_s)^3+\frac{2\pi}{3} R_s^3 & z_s - R_s \le z_{sf} \le z_s + R_s \\
	 \frac{4}{3} \pi R_s^3 & z_{sf} > z_s + R_s
	\end{array} \right.  \label{eq:volume}  \\
V_3 & = & \pi R_{cyl}^2 z_{cf}, \nonumber
\end{eqnarray}
with $V_{bub} = V_3 - V_2 - V_1$
The first term ($V_1$) is the volume of revolution about the $z$ axis of the film between its endpoints, and the second term ($V_2$) is the volume of the spherical cap below the the point of contact between the film and the sphere. These are both subtracted from the third term ($V_3$), which is the total cylindrical volume enclosed by the outer wall of the cylinder beneath the point of contact $z_{cf}$ between the film and the cylinder wall.

\subsection{Forces}

We consider two forces in addition to the weight $mg$ acting in the negative $z$ direction. The tension force $\underline{F}_\gamma$ is due to the pull of the soap film around its circular line of contact with the sphere and the pressure force $\underline{F}_p$, which is only relevant in case 1, is due to the pressure in the trapped bubble which acts over the surface of the sphere below the contact line. We are interested only in the vertical component of these forces, since by symmetry the other components cancel.

We define the angle $\theta$ that the film subtends with the centre of the sphere, $\tan \theta = (z_{sf}-z_s)/r_{sf}$, and then the $z-$components of the forces are
\begin{equation}
 F_{\gamma} = 2 \gamma . \; 2 \pi r_{sf} \; \cos (\theta - \theta_c) 
\end{equation}
and
\begin{equation}
 F_p =  \pi r_{sf}^2 \; p_{bub}, 
\label{eq:press_force}
\end{equation}
where $p_{bub}$ is the pressure in the bubble.
 
\subsection{Motion}

We perform a quasi-static simulation in which the position of the sphere is held fixed while the equilibrium shape of the film is found, and then the sphere is moved a small distance in the direction of the resultant force. In case 1 the bubble pressure is found from the Lagrange multiplier of the volume constraint, eq.~(\ref{eq:volume}). 

\begin{figure}
\centerline{
(a)
 \includegraphics[angle=0,width=0.4\textwidth]{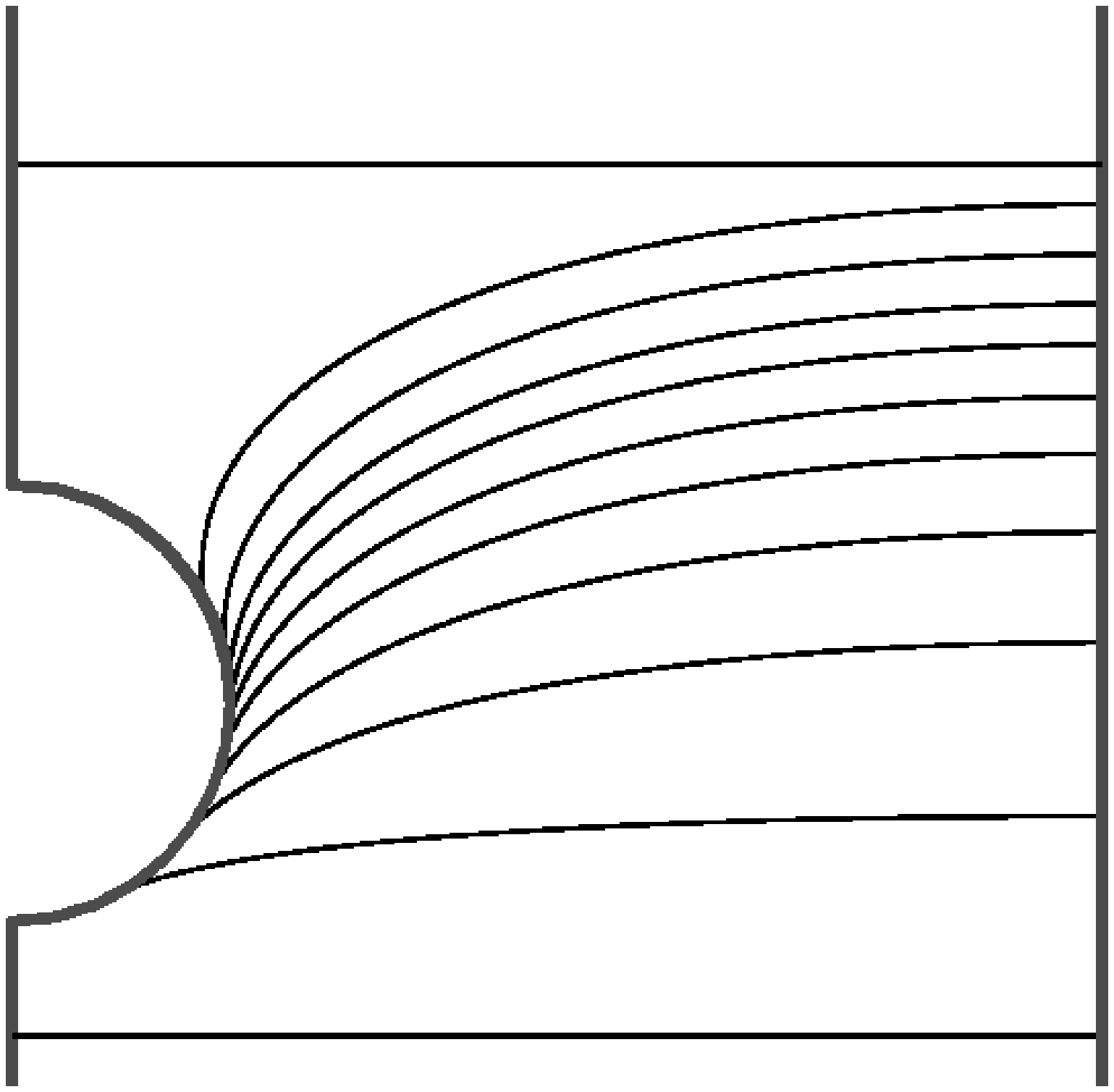}
(b)
 \includegraphics[angle=0,width=0.4\textwidth]{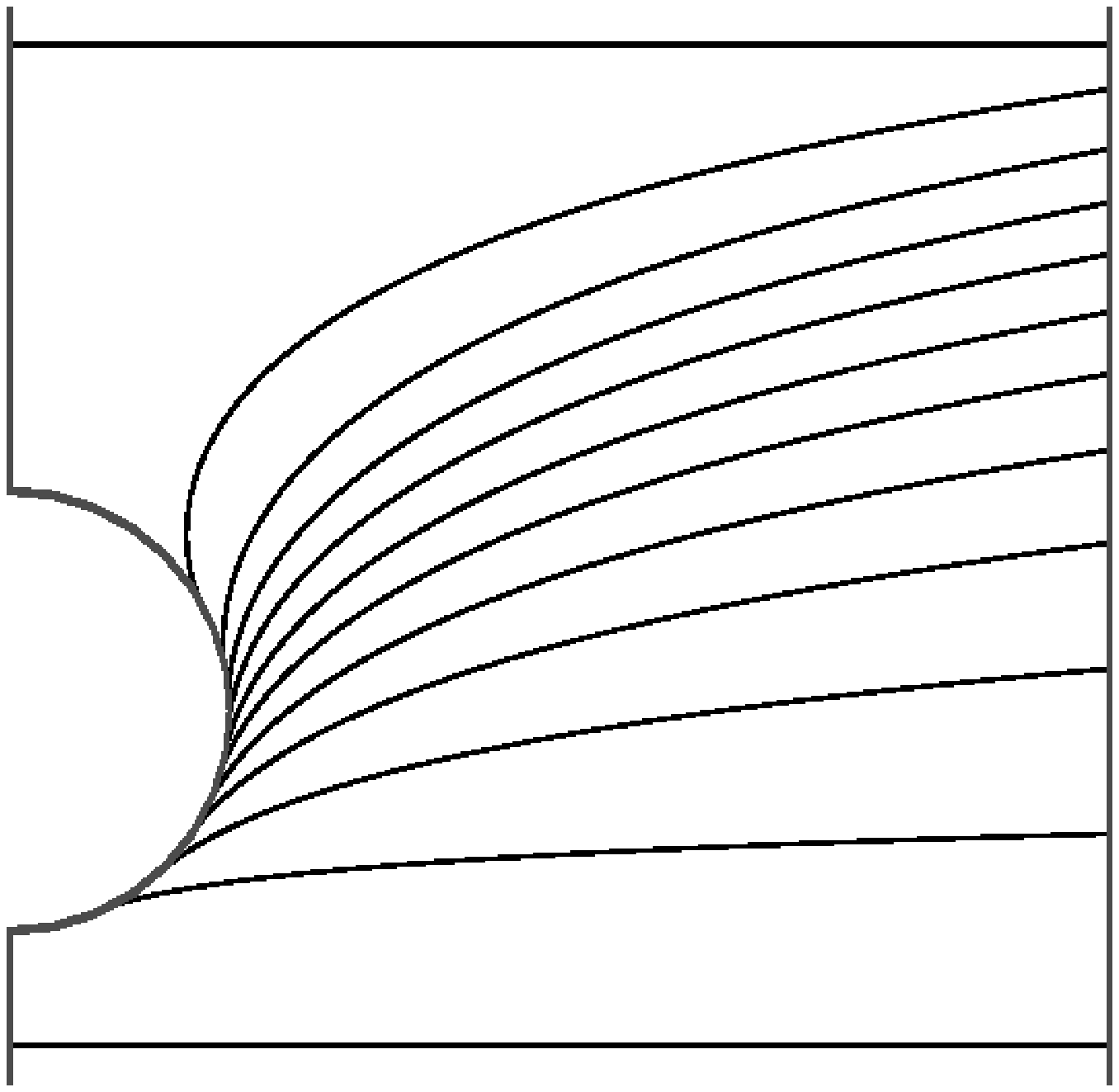}
  }
\caption{Film shapes in a frame of reference moving with the sphere, with contact angle $\theta_c = 10^\circ$, shown every 100 iterations. (a) Case 1, where a wetting film on the outer cylinder wall allows the film to slip there and hence meet the wall at $90^\circ$. (b) Case 2, where the film is fixed at the outer cylinder wall. }
\label{fig:film_shape}
\end{figure}

We start the simulation with the sphere above a horizontal film, and move the sphere downwards in steps of $\Delta z_s = - \epsilon mg$, with the small parameter $\epsilon$ taken equal to  $1\times 10^{-5}$ (which we find is sufficiently small not to change the results), until contact is made. The inner end of the film then jumps to a new position on the sphere and then the change in its vertical position obeys
\begin{equation}
\Delta z_s = \epsilon \left( F_\gamma + F_p - mg \right).
\end{equation}
Detachment occurs when the film nears the top of the sphere and becomes unstable, at which point it jumps back to being horizontal, and we then end the simulation. Note that $\Delta z_s$ is always negative in our simulations, since the weight of the sphere is large enough that it always exceeds the tension force. 

\section{Results}
\label{sec:results}

\begin{figure}
(a)
\centerline{
 \includegraphics[angle=270,width=0.5\textwidth]{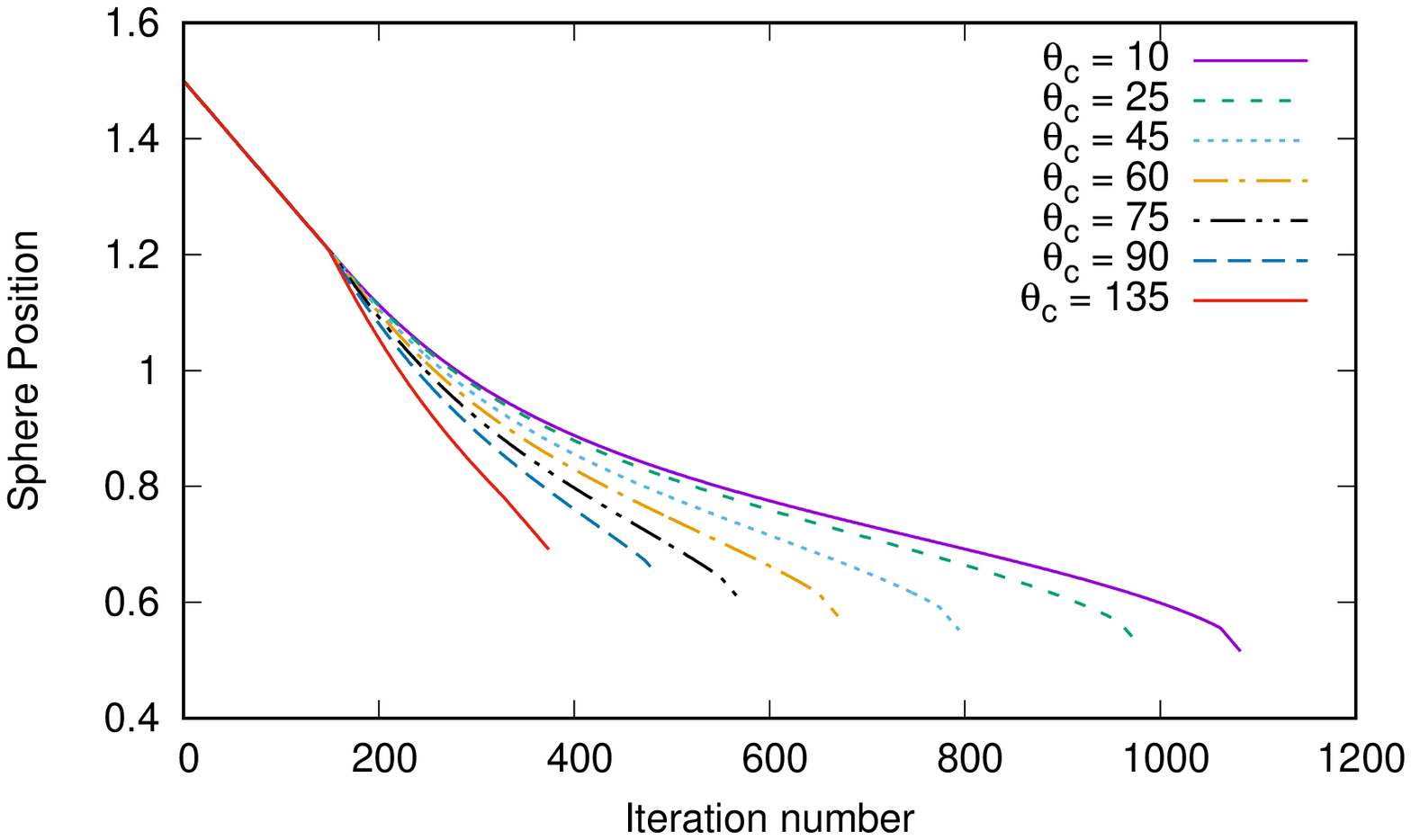}
(b)
 \includegraphics[angle=270,width=0.5\textwidth]{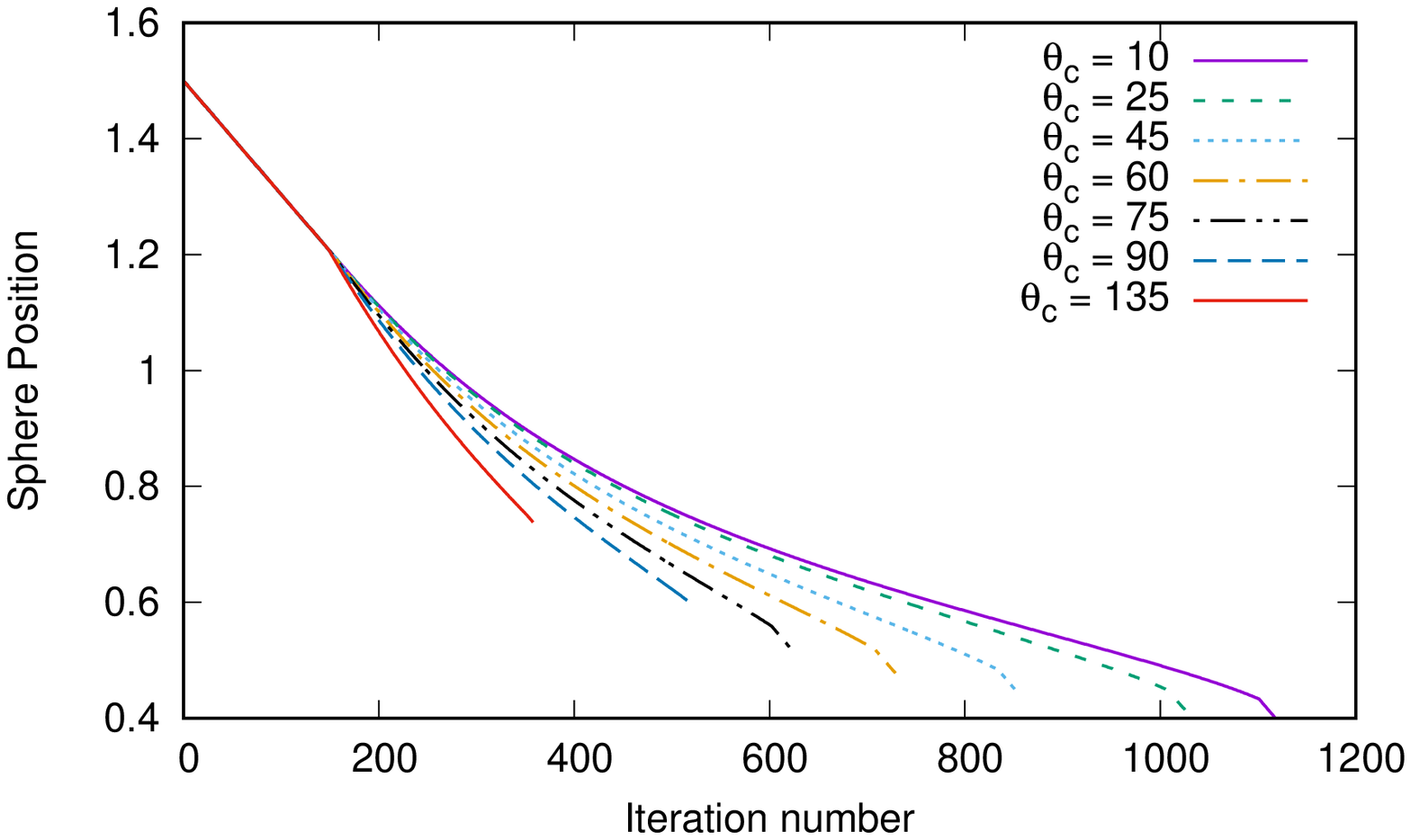}
}
\caption{The height of the centre of the sphere under the action of its weight and the forces that the foam exerts on it. The horizontal axis corresponds to time, in units of $\epsilon$. (a) Case 1. (b) Case 2.}
\label{fig:centre_of_sphere}
\end{figure}

In this section we consider a sphere of radius $R_s = 0.2{\rm cm}$ and mass $m = 0.1$ grams. Then the particle density is $\rho \approx 3 g/{\rm cm}^3$ and the Bond number is $Bo \approx 2 $. An example of the shape of the film at different times is shown in figure \ref{fig:film_shape}.

\subsection{Sphere position, soap film area, and point of contact}

The vertical position of the centre of the sphere is shown in figure~\ref{fig:centre_of_sphere}. Before attachment the sphere follows the same path for all contact angles. Following attachment (at an iteration number close to 150) we observe a shallower curve for smaller contact angles, indicating that the forces retard the motion of the sphere to a greater extent when the contact angle is small. When the contact angle is larger, for example with $\theta_c$ greater than about $45^\circ$, the sphere motion is at first accelerated, as the film pulls it downwards. In case 1, the bubble pressure is also negative at first (see figure~\ref{fig:pressure_force} below), adding to this effect. For the contact angle of $\theta_c = 135^\circ$ this significantly reduces the time of interaction before the film detaches from the top of the sphere. 

After detachment the slope of each curve returns to the same value as before attachment (data not shown) for all contact angles. In case 2, without a volume constraint, the interaction time (when the film and sphere are in contact) is longer for each value of contact angle compared to case 1, and the sphere descends further before detachment. Hence the overall effect of constraining the volume rather than the outer rim of the film is to retard the sphere.

\begin{figure}
(a)
\centerline{
 \includegraphics[angle=270,width=0.5\textwidth]{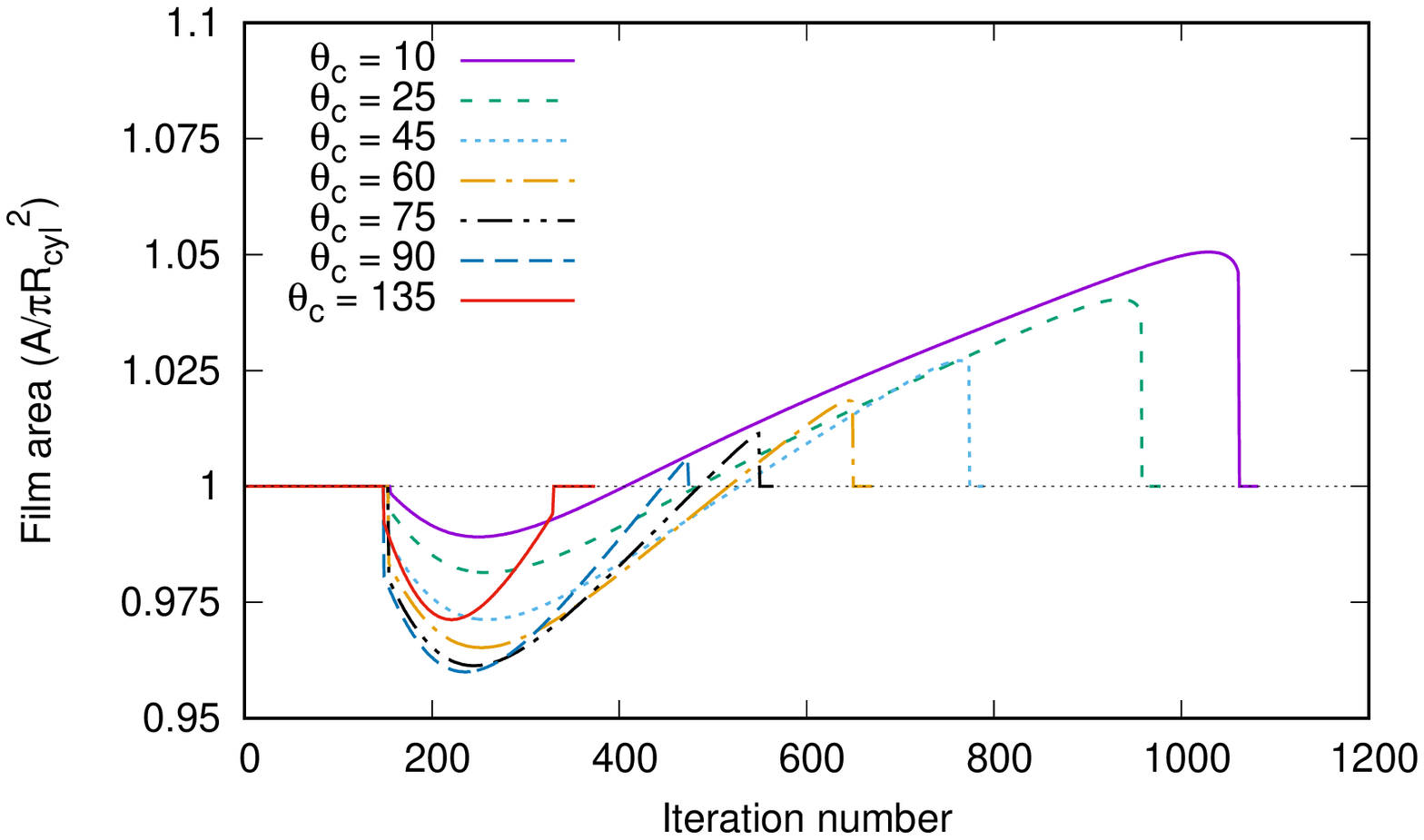}
(b)
 \includegraphics[angle=270,width=0.5\textwidth]{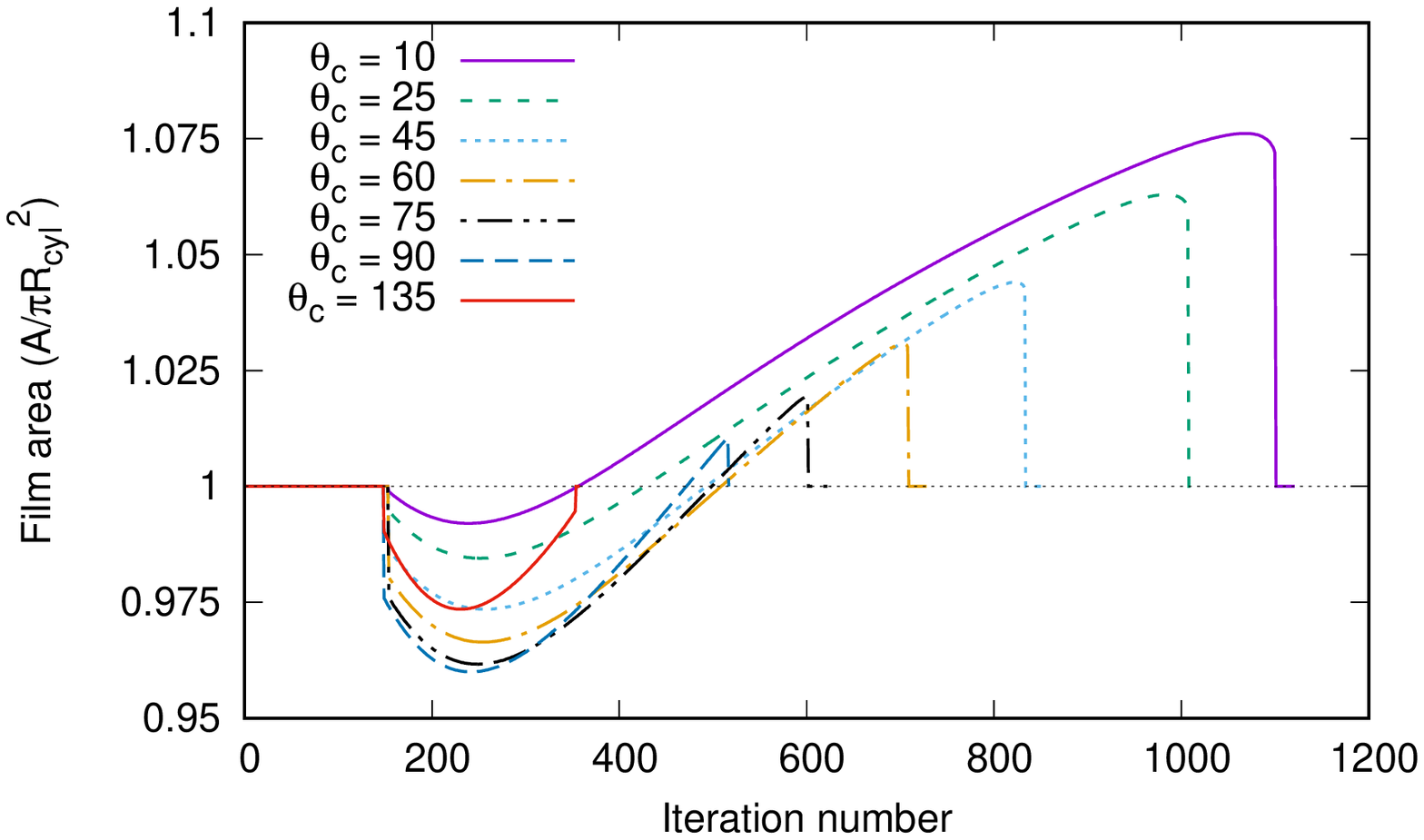}
}
\caption{The area of the soap film as the sphere passes through it. (a) Case 1. (b) Case 2. }
\label{fig:area_of_film}
\end{figure}

Detachment occurs {\em before} the inner end of the soap film reaches the top of the sphere. Instead, there is a sort of  ``pre-emptive" instability~\cite{Hutzlerwcve07}: the curved soap film becomes unstable, the line of contact jumps upwards, and a new configuration consisting of a flat film above the sphere is reached. This is seen, for example, in the abrupt jump in the surface area of the film, shown in figure~\ref{fig:area_of_film}, at the point of detachment. 

When the sphere first meets the film the film area is reduced because it contains a circular hole that is filled by the sphere. As the sphere descends further, the film is deformed, in order to obey the volume constraint (in case 1) or the fixed rim at the cylinder wall (in case 2) and to satisfy the contact angle where they meet. This causes the film area to increase, until the film approaches the point of detachment. For a contact angle of $135^\circ$ (and presumably greater) the area of the film never exceeds its equilibrium value, $A=\pi R_{cyl}^2$, indicating that it is not greatly deformed and that detachment occurs quickly. Comparing case 1 to case 2, for all other contact angles simulated, the film is slightly more deformed when its outer rim is fixed (case 2).

\begin{figure}
(a)
\centerline{
 \includegraphics[angle=270,width=0.5\textwidth]{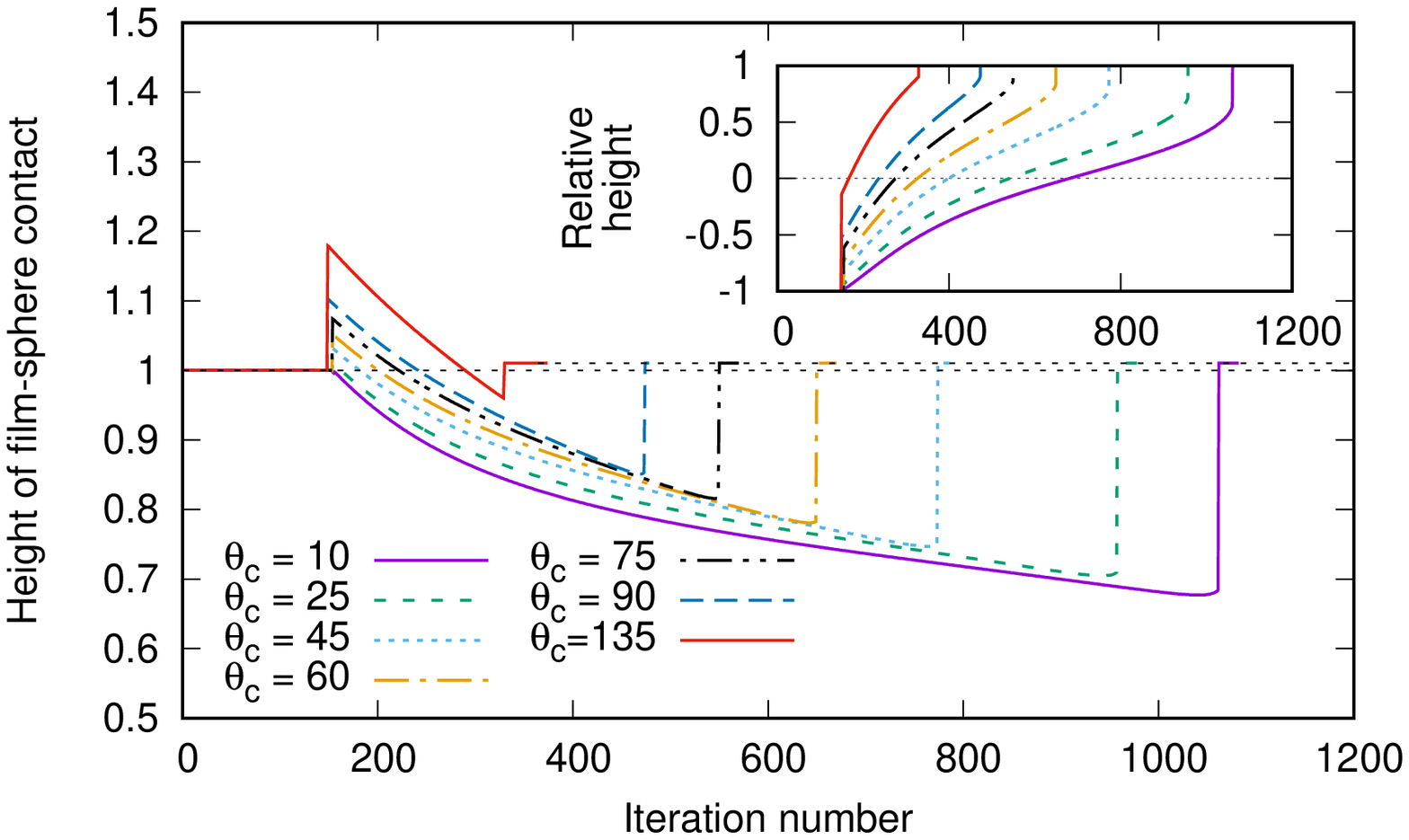}
(b)
 \includegraphics[angle=270,width=0.5\textwidth]{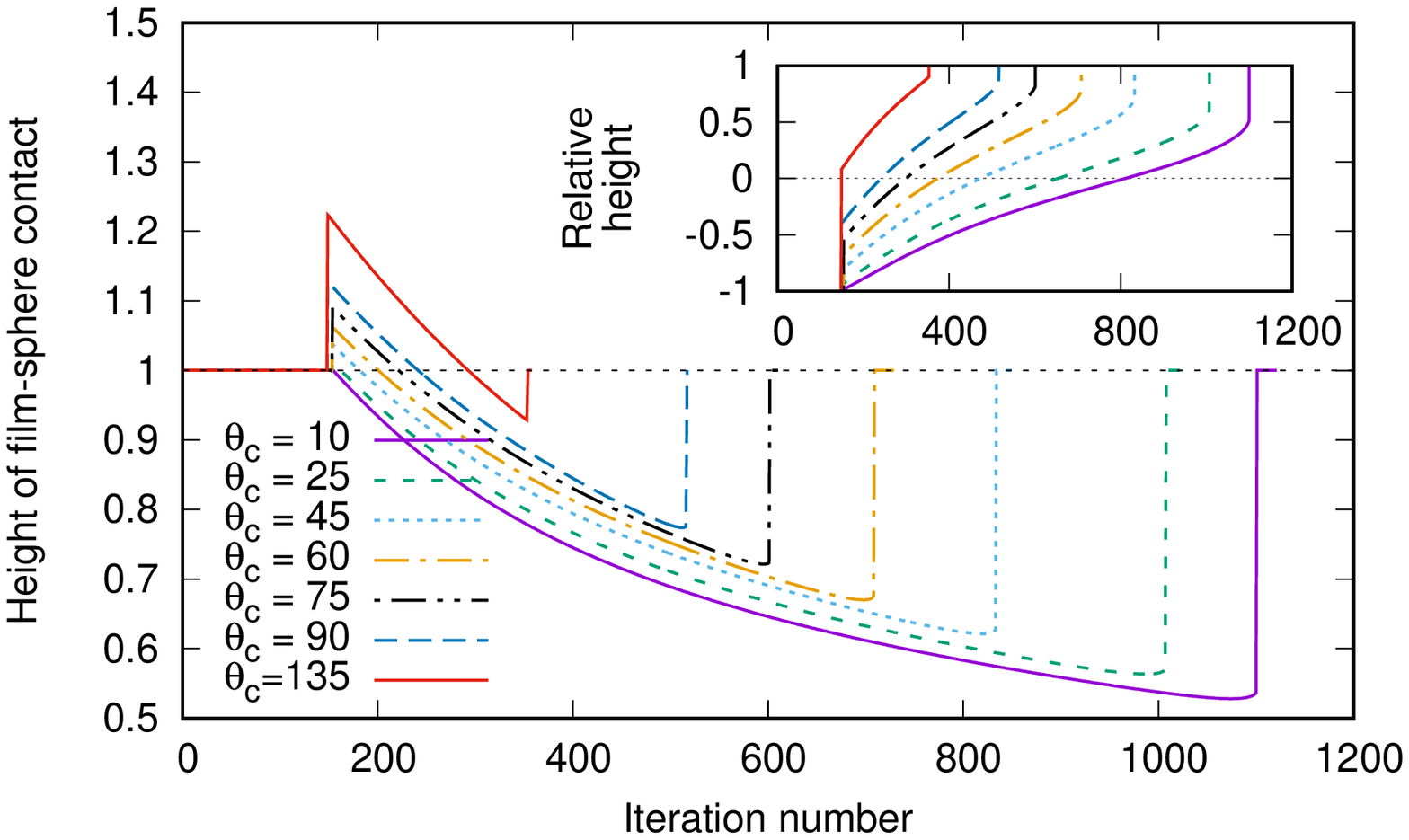}
}
\caption{The vertical position $z_{sf}$ of the line where the film touches the sphere. The inset shows this position relative to the height of the centre of the sphere, $(z_{sf}-z_s)/R_s$. (a) Case 1. (b) Case 2.}
\label{fig:vertex_position}
\end{figure}

Just as there is a sudden jump during detachment, there is also a jump in the vertical position of the circular line of contact when the film first meets the sphere (figure~\ref{fig:vertex_position}). The contact line rises to a new position to satisfy the contact angle (without, in case 1, violating the volume constraint), to a degree that increases with the contact angle. This end of the film is then pulled down by the sphere, more so for large contact angles, and the decrease is monotonic until detachment, whereupon the film is suddenly released. In case 1, the film returns to a higher position after the sphere has passed, because the volume enclosed beneath the film is augmented by the volume of the sphere. 

Fixing the outer rim of the film (case 2) leads to a greater deformation of the film (figure~\ref{fig:area_of_film}) and hence to the film becoming unstable when the line of contact is further from the top of the sphere (figure~\ref{fig:vertex_position} insets).

In case 1 the outer rim of the film, where it touches the cylinder wall, behaves slightly differently (data not shown). It at first drops suddenly, i.e. in the opposite sense to the inner contact line, and then increases until the inner contact line approaches the top of the sphere. It then descends again before suddenly returning to the same vertical position as the inner contact line when the film detaches and becomes flat.

\subsection{Measured forces}

\begin{figure}
(a)
\centerline{
 \includegraphics[angle=270,width=0.5\textwidth]{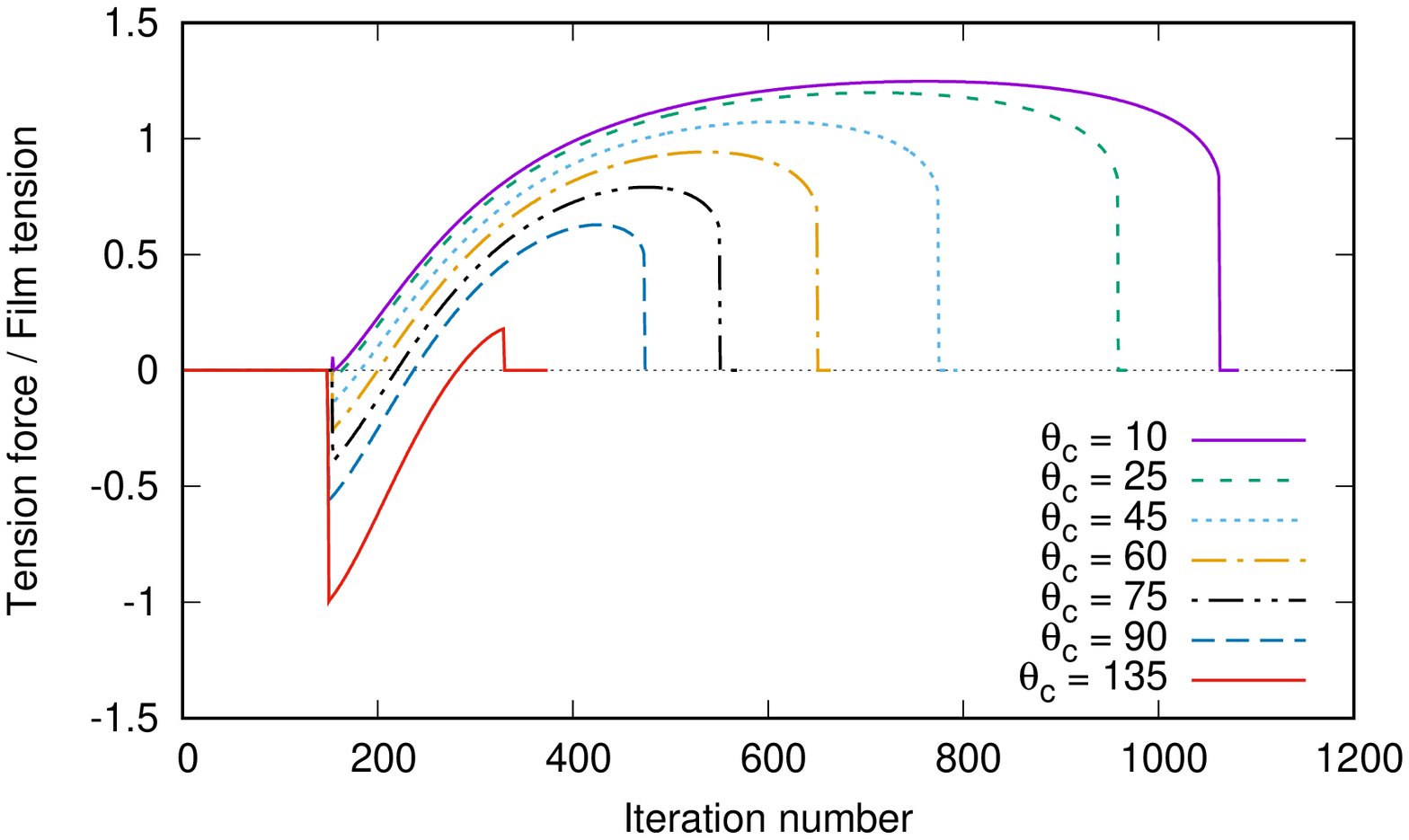}
(b)
 \includegraphics[angle=270,width=0.5\textwidth]{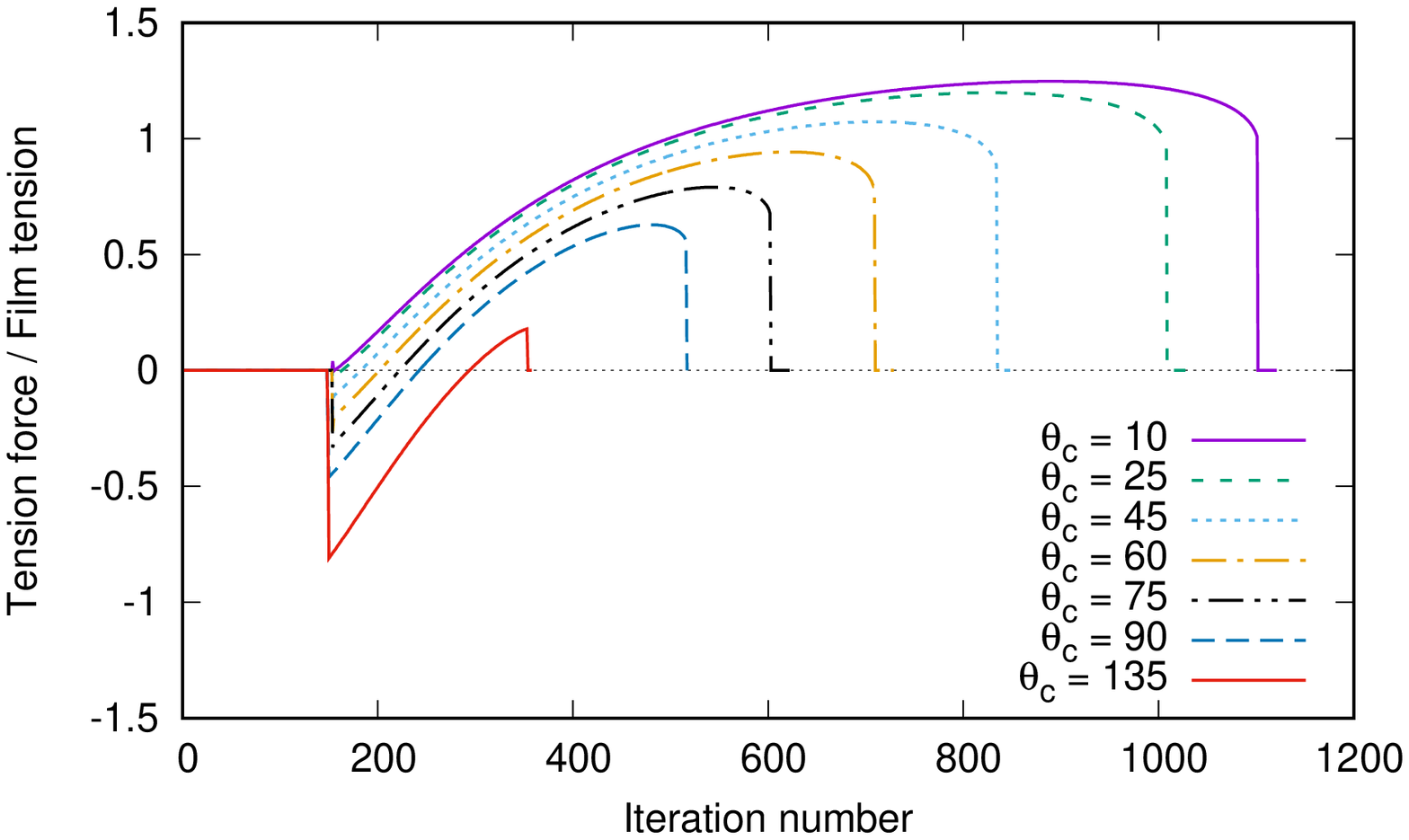}
}
\caption{Tension forces exerted on the sphere, determined by the direction in which the film pulls multiplied by its tension. (a) Case 1. (b) Case 2. }
\label{fig:tension_force}
\end{figure}

\begin{figure}
\centerline{
 \includegraphics[angle=270,width=0.5\textwidth]{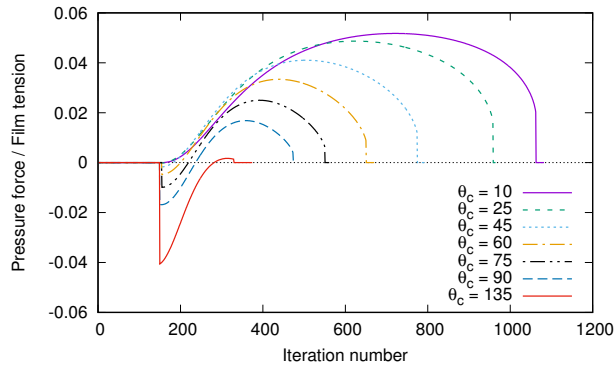}
  }
\caption{Pressure forces exerted on the sphere in Case 1.}
\label{fig:pressure_force}
\end{figure}

We show the forces acting on the sphere in figures~\ref{fig:tension_force} and ~\ref{fig:pressure_force}. For large contact angles the film pulls the sphere downwards, accelerating its motion. The opposite occurs for small contact angles, and so the time over which the sphere contacts the sphere is extended. Just before the abrupt drop in force at the point of detachment, there is a slight reduction in the tension force as the perimeter of the contact line becomes small, ameliorating the pull from the film. 

In case 1, the pressure in the bubble can be either positive or negative, depending on the curvature of the film. The pressure force on the sphere is determined by this pressure multiplied by the vertically-projected area of the sphere over which the bubble touches the sphere, eq. (\ref{eq:press_force}). The pressure force is much smaller in magnitude than the tension force. For the contact angle of $135^\circ$ the bubble pressure is large and negative for much of the passage of the sphere, because of the curvature induced by the contact angle, so in this case the pressure force ``sucks" the sphere downwards and detachment occurs earlier than in case 2. 

For smaller contact angles, for example $\theta_c=10^\circ$, the pressure is always positive, opposing the downward motion of the sphere. Yet it is still the case that detachment occurs sooner in case 1, even though for a given contact position the tension force is similar in both cases. Further, the film becomes unstable at a lower position in case 2. The resolution of this apparent paradox is that when the contact line is at a certain position on the sphere, the sphere is at a different height in the two cases, because of the need to satisfy the different constraints and for the film to meet the sphere at the same contact angle. In particular, before the contact line passes the equator of the sphere ($z_{sf} < z_s$), it moves around the sphere more slowly in case 2, while above the equator it moves more quickly (but over a shorter distance).

\section{Bubble entrainment}

\begin{figure}
(a)
\centerline{
 \includegraphics[angle=0,width=0.3\textwidth]{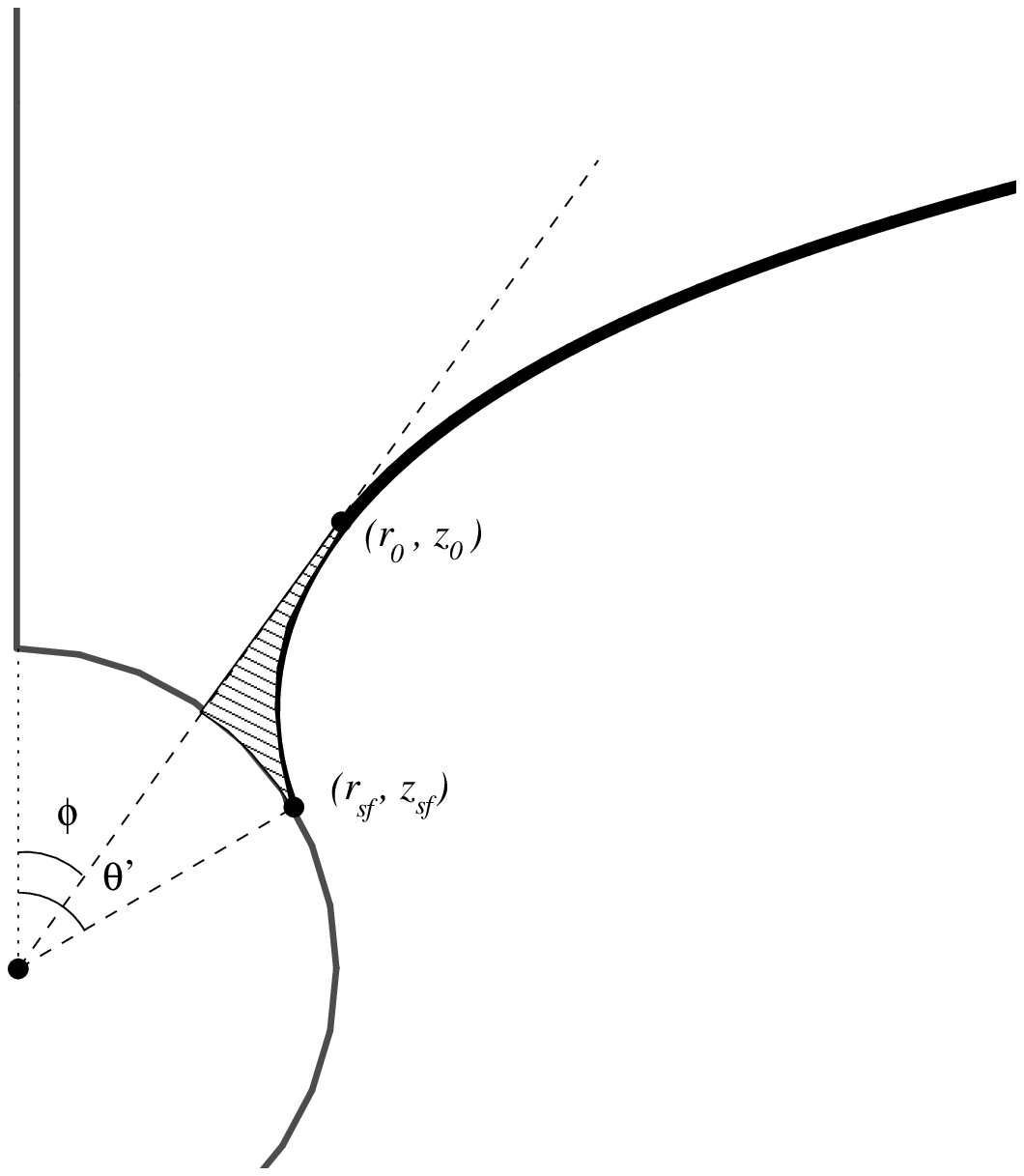}
(b)
 \includegraphics[angle=0,width=0.5\textwidth]{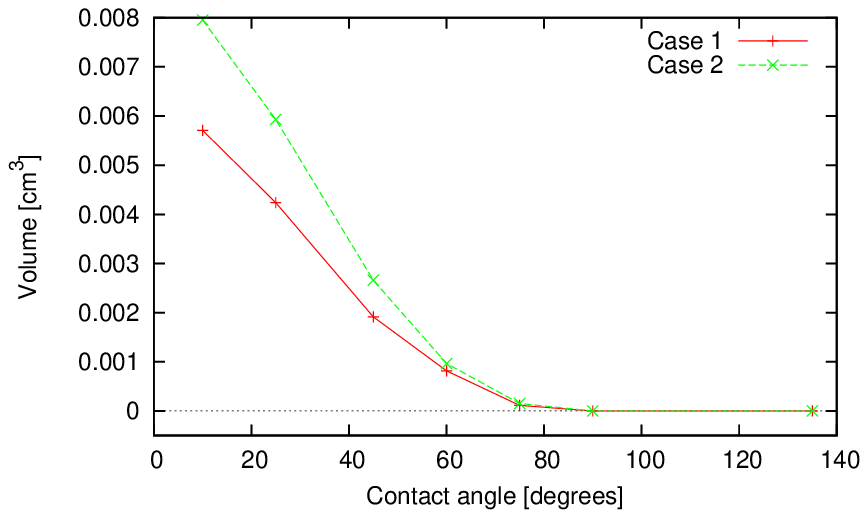}
  }
\caption{(a) Close to the contact line between the soap film and the sphere, at the last iteration before detachment, we calculate the shaded area to estimate the volume of the small bubble that is left behind. (b) The bubble volume depends strongly on the contact angle, depends only weakly on whether we consider case 1 or case 2, and vanishes for contact angles greater than $90^\circ$.}
\label{fig:satellite}
\end{figure}

Although our quasistatic simulations do not resolve the rapid film motion during detachment, we can gain an idea of the size of the small bubble that is trapped~\cite{legoffcsq08} by examining the shape of the soap film immediately before detachment, as shown in figure~\ref{fig:satellite}. We calculate the area of the region in the $(r, z)$ plane that is shaded in the figure, between the soap film and a radial line through the point of the soap film closest to the vertical axis, and rotate this region about the vertical axis to estimate the bubble volume. This is likely to be an underestimate, as the curvature of the film around the catenoidal neck is likely to increase during detachment.


Figure~\ref{fig:satellite} shows that for small contact angles the bubble size can reach almost $0.01 {\rm cm}^3$. The limit in which the contact angle tends to zero appears to give a well-defined value for the maximum size of this small satellite bubble. For contact angles of $90^\circ$ and above there is no bubble because the point on the soap film nearest to the vertical axis is where the film touches the particle.

There is a small effect of the choice of boundary conditions: in case 2, without a pressure force, the bubble is about 30\% larger for $\theta = 10^\circ$ (although this difference decreases as the contact angle increases). This is because, as noted above, in case 2 the instability that causes the film to detach occurs earlier, when the line of contact is closer to the equator of the sphere. 

\begin{figure}
\centerline{
 \includegraphics[angle=0,width=0.5\textwidth]{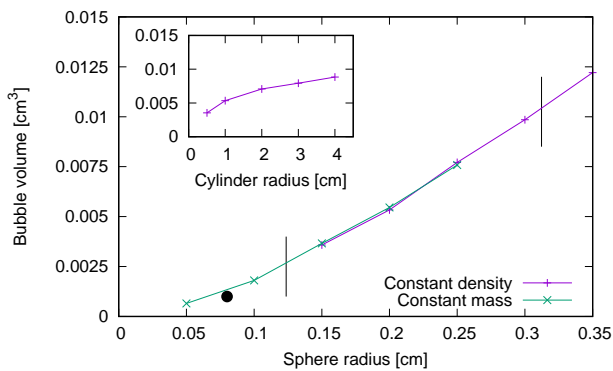}
  }
\caption{With contact angle $\theta_c = 10^\circ$ in case 1, the volume of the bubble that is trapped by the film increases with the size of the particle and (inset) depends weakly on the size of the cylinder containing the soap film. With fixed mass $m=0.12g$ only spheres with radius up to $R_s = 0.25 cm$ pass through the film; with constant density $\rho \approx 6 {\rm g/cm}^3$ only spheres with radius larger than $R_s = 0.10 cm$ pass through the film; the size of the trapped bubble is the same in both cases, indicating that it is determined by the geometry of the soap film. The vertical lines indicate the radius bounds estimated at the end of \S \ref{sec:intro} and the solid circle is experimental data~\protect\cite{legoffcsq08}.}
\label{fig:satellite2}
\end{figure}

In case 1 with a fixed contact angle of $10^\circ$ we varied the size of the spherical particle and again estimated the size of the trapped bubble. For a sphere of a given radius, we must choose between a fixed particle mass (weight) or a fixed particle density. In the former case, the tension force opposing the descent of the particle increases with particle radius, but since the sphere does not increase in weight, it is brought to rest by the soap film once the particle exceeds a critical radius (in our case with $R_s \approx 0.3$cm, or $Bo \approx 3$). In the latter case, only when the particle is sufficiently large (in our case with $R_s$ greater than 0.1cm) does it pass through the soap film. Figure~\ref{fig:satellite2} shows that the size of the bubble that is trapped is the same in both cases. So it is determined by the shape of the soap film only, which in turn arises from the film meeting the sphere, of whatever radius, at the given contact angle. Therefore the size of the trapped bubble increases with sphere size, since the film is more greatly deformed when the sphere is larger. 

There is also a small dependence of the size of the trapped bubble on the cylinder size. As the cylinder becomes larger, the sphere descends further before detachment, and so greater film deformation is possible. In addition, the pressure force is reduced in a larger cylinder, so the result should be closer to case 2. Thus, the trapped bubble is slightly larger if the cylinder radius is larger.

Figure~\ref{fig:satellite2} also shows the minimim and maximum sphere radius for which the sphere passes through the film, based on the predictions in \S \ref{sec:intro}. The lower bound (for constant particle density) is just below the numerical data while the upper bound (for constant particle mass) is about 20\% above the upper limit of the data in that case. The bounds are predicated on the soap film pulling vertically upwards around the equator of the sphere, but despite this approximation appear to work well.

To validate our predictions, we compare with the image in Figure 1 of~\cite{legoffcsq08}, which shows a sphere of radius 0.16cm falling through a soap film trapping a bubble. (The cylinder radius and sphere mass are not recorded.) The bubble is trapped against the upper part of the sphere, but appears to be roughly hemispherical with radius 0.08cm, and hence a volume of 0.001cm$^3$. The data point, shown in figure~\ref{fig:satellite2}, lies close to our prediction.

\section{Conclusions}

We have explained the effect of contact angle on the forces that act on a spherical particle passing through a soap film. The duration of the interaction is determined by the contact angle and also the way in which the film is deformed; for example, with low contact angles the particle moves more slowly, and stays in contact with the soap film for longer. Further, the interaction depends upon the details of the experiment: greater deformation is induced by holding the film in a fixed circular wire frame than in a cylindrical tube, where it traps a bubble but where the outer circumference of the film is not fixed, such as in a soap-film meter \cite{chenpjhd19}. In the latter case there is an additional force on the particle due to the pressure in the bubble, but this is negligible in determining the dynamics of the system.

Analysing the shape of the soap film just before detachment allows us to predict the size of the small bubble that is formed when a particle passes through a film. The entrapment of this air and the formation of interface could play a role in determining the efficacy of using foams for the suppression of explosions. We find that the bubble increases in size as the particle gets larger, and can exceed $10 {\rm mm}^3$.

Extending our predictions to more general cases, such as oblique impact and non-spherical particles \cite{morrisnc12,davies18}, will require more computationally-intensive three-dimensional simulations.

\section*{Acknowledgements}

The late J.F. Davidson inspired us to work on this problem.
We are also grateful to C. Raufaste for useful discussions, and to K. Brakke for provision and support of the Surface Evolver software. 
SJC acknowledges financial support from the UK Engineering and Physical Sciences Research Council (EP/N002326/1).


\end{document}